\documentclass[11pt,a4paper]{article}
\usepackage{geometry}
\usepackage[english]{babel}
\usepackage{amsmath,amsthm}
\usepackage{amsfonts}
\usepackage{graphicx}
\usepackage{bm}
\usepackage[colorlinks,linkcolor=blue,anchorcolor=blue,citecolor=blue,urlcolor=blue,filecolor=blue,menucolor=blue,runcolor=blue]{hyperref}
\usepackage[square, comma, sort&compress, numbers]{natbib}
\usepackage{ulem}
\usepackage{authblk}

\begin{document}

\title{Localized 4f-electrons in the quantum critical heavy fermion ferromagnet CeRh$_6$Ge$_4$}

\author[1]{An Wang}
\author[1]{Feng Du}
\author[1,2]{Yongjun Zhang}
\author[3]{David Graf}
\author[1]{Bin Shen}
\author[1]{Ye Chen}
\author[1,4]{Yang Liu}
\author[1,4]{Michael Smidman}
\author[5,1]{Chao Cao}   
\author[1,6]{Frank Steglich}   
\author[1,4,7]{Huiqiu Yuan \thanks{Corresponding author: hqyuan@zju.edu.cn}}

\affil[1] {\textit{Center for Correlated Matter and Department of Physics, Zhejiang University, Hangzhou 310058, China}}
\affil[2] {\textit{Institute for Advanced Materials, Hubei Normal University, Huangshi 435002, China}}
\affil[3] {\textit{National High Magnetic Field Laboratory and Department of Physics, Florida State University, Tallahassee, Florida 32306, USA}}
\affil[4] {\textit{Zhejiang Province Key Laboratory of Quantum Technology and Device, Department of Physics, Zhejiang University, Hangzhou  310058, China}}
\affil[5] {\textit{Department of Physics, Hangzhou Normal University, Hangzhou 310036, China}}
\affil[6] {\textit{Max Planck Institute for Chemical Physics of Solids, Dresden, Germany}}
\affil[7] {\textit{State Key Laboratory of Silicon Materials, Zhejiang University, Hangzhou 310058, China}}

\date{}
						

\maketitle
\textbf{Abstract}

Ferromagnetic quantum critical points were predicted to be prohibited in clean itinerant ferromagnetic systems, yet such a phenomenon was recently revealed in CeRh$_6$Ge$_4$, where the Curie temperature can be continuously suppressed to zero under a moderate hydrostatic pressure. Here we report the observation of quantum oscillations in CeRh$_6$Ge$_4$ from measurements using the cantilever and tunnel-diode oscillator methods in fields up to 45~T, clearly demonstrating that the ferromagnetic quantum criticality occurs in a clean system.  In order to map the Fermi surface of CeRh$_6$Ge$_4$, we performed angle-dependent measurements of quantum oscillations at ambient pressure, and compared the results to density functional theory calculations. The results are consistent with the Ce 4f electrons remaining localized, and not contributing to the Fermi surface, suggesting that localized ferromagnetism is a key factor for the occurrence of a ferromagnetic quantum critical point in CeRh$_6$Ge$_4$.\\


\textbf{1.Introduction}

The continuous suppression of a second-order phase transition to zero temperature at a quantum critical point (QCP) profoundly affects the finite temperature physical properties, leading to the breakdown of Fermi-liquid behavior, and the occurrence of novel ordered phases  such as unconventional superconductivity \cite{Stewart2001, Gegenwart2008, Weng2016, Michael2019, ChenYe2016}. Quantum criticality upon suppressing ferromagnetic (FM) order is however ordinarily not observed; upon tuning the system the ferromagnetism either abruptly disappears at a first-order transition, or there is a change of magnetic ground state \cite{Brando2016, Pfleiderer1997, Sullow1999, Brando2008, Pfleiderer2002}. These experimental observations are consistent with the theoretical prediction that soft fermionic modes prevent the occurrence of FM QCP's in clean  itinerant FM systems, which are instead driven first-order \cite{Belitz1999,Chubukov2004}. Although FM quantum criticality has been reported to be induced by chemical substitution in some materials \cite{Steppke2013,Sokolov2006,Jia2011,Huang2020}, the non-stoichiometric nature leaves open the predicted scenario of a disorder-driven continuous FM quantum phase transition.

Very recently stoichiometric CeRh$_6$Ge$_4$, a heavy fermion ferromagnet with $T_{\rm C}=$2.5~K \cite{Matsuoka2015}, was  found to exhibit a pressure-induced FM QCP at $p_c=0.8$ GPa, which is accompanied by a strange metal phase exhibiting a remarkable resemblance to the  high-$T_{\rm c}$ cuprate superconductors, with a linear in temperature resistivity and a logarithmic divergence of the specific heat coefficient \cite{Shen2019}. These findings were reconciled with the predicted prohibition of FM QCP's in clean \textit{itinerant} systems by the proposal of \textit{local} quantum criticality \cite{Shen2019}. Here  there is a critical destruction of the Kondo effect at zero-temperature   (Kondo breakdown) and localization of the Ce $4f$-electron moments at $p_c$, upon approaching from the heavy Fermi liquid region, leading to an abrupt decrease of the Fermi surface volume at the QCP \cite{Si2001,Coleman2001}. A similar scenario has been proposed to occur in a handful of AFM heavy fermion systems such as CeCu$_{6-x}$Au$_x$, YbRh$_2$Si$_2$ and CeRhIn$_5$ \cite{Schroder2000,Custers2003,Paschen2004,Shishido2005}. Alternatively, the FM QCP was attributed to  `quenched disorder', arising from strong  antisymmetric spin-orbit coupling (ASOC) due to broken inversion symmetry, which leads to the soft modes, which would otherwise inhibit FM quantum criticality, acquiring mass \cite{Kirkpatrick2020}. 

In order to scrutinize the origin of the FM QCP in CeRh$_6$Ge$_4$ and understand the physical properties, it is vital to characterize the electronic structure of  CeRh$_6$Ge$_4$. In this article we report angle dependent measurements of quantum oscillations in high quality single crystals of CeRh$_6$Ge$_4$ at ambient pressure, in order to map the Fermi surface and to compare the results to density functional theory (DFT) calculations. Our observation of quantum oscillations demonstrates that the previously observed  QCP in CeRh$_6$Ge$_4$ occurs in a clean FM system \cite{Shen2019}, ruling out the scenario of a disorder-driven suppression of a first-order transition \cite{Sang2014}. The comparison of the results with DFT calculations provides evidence that the Ce-$4f$ moments  of CeRh$_6$Ge$_4$ are localized at ambient pressure. This is distinct from itinerant ferromagnets where FM QCP's are not observed \cite{Brando2016}, and is  in accordance with the Kondo breakdown scenario.\\

\textbf{2. Methods}

\begin{figure} [tb]
	\includegraphics[angle=0,width=0.8\textwidth]{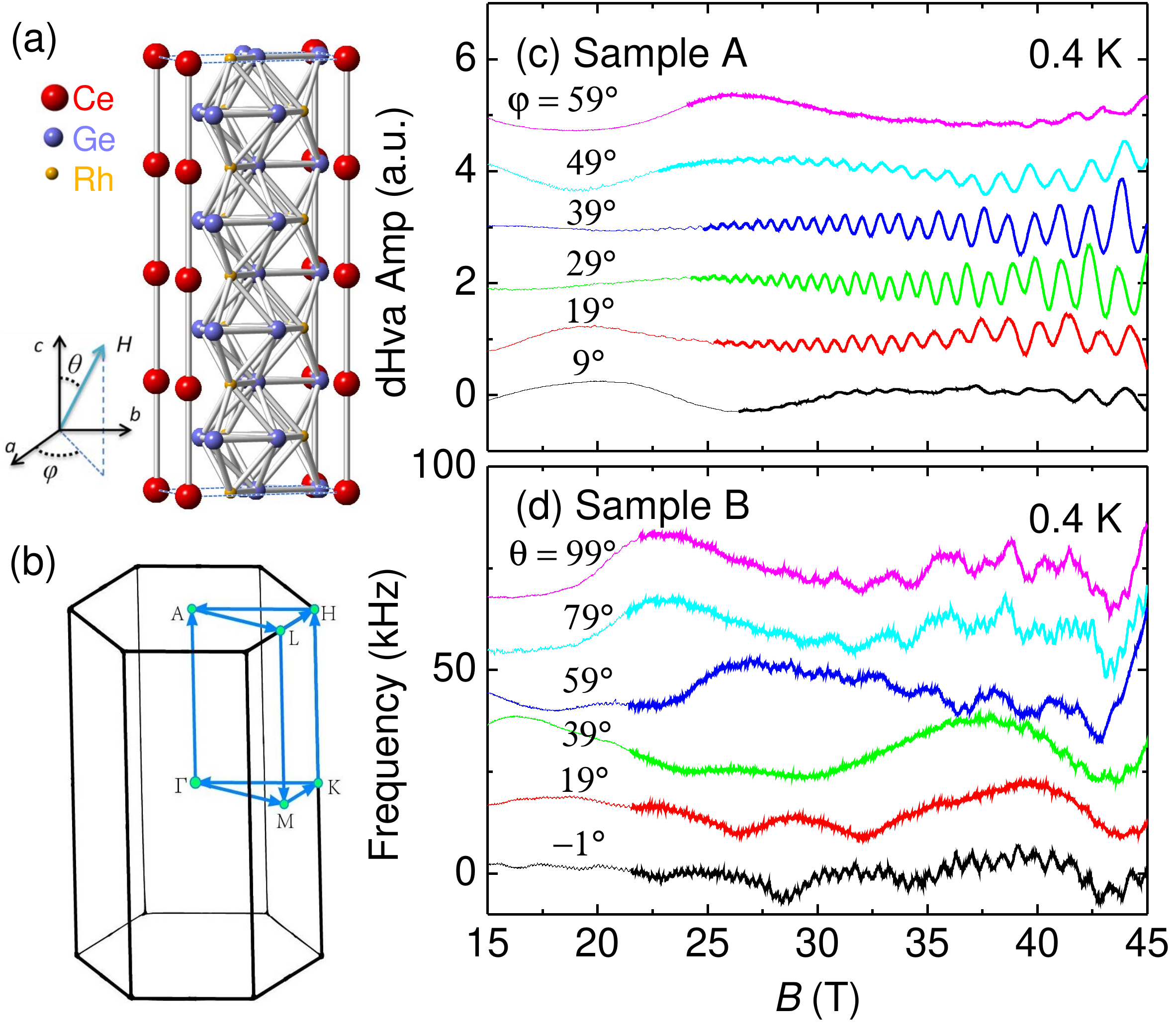}
	\centering
	\vspace{-10pt}
	\caption{\label{setup}  (Color online) (a) Crystal structure, and (b) Brillouin zone of CeRh$_6$Ge$_4$. Red, blue and yellow atoms correspond to Ce, Rh and Ge respectively. Quantum oscillations at 0.4~K after subtracting the background are displayed for (c) sample A for different field-angles $\varphi$ within the $ab$-plane, and (d) sample B for different field-angles $\theta$ to the $c$-axis.  }
\end{figure}

Single crystals of CeRh$_6$Ge$_4$  were grown using a Bi flux \cite{Shen2019,164growth}. The hexagonal crystal structure (space group $P\bar{6}m2$) and diagram of the Brillouin zone are shown in Figs.~\ref{setup}(a) and (b), respectively. Measurements in high-magnetic fields up to 45~T were performed at the National High Magnetic Field Laboratory, Tallahassee, using the torque-magnetometer method for  fields rotated within the $ab$-plane (sample A), with a fixed polar angle $\theta=90^{\circ}$, and the tunnel-diode oscillator method  for  fields applied between the $ab$-plane and $c$-axis (sample B) at a fixed azimuthal angle $\varphi=30^{\circ}$. DFT calculations were  implemented in the VASP software, including the effect of spin-orbit coupling using the second variational method.  Typical quantum oscillations are  displayed in Figs.~\ref{setup}(c) and (d) upon varying $\theta$ and $\varphi$, after subtracting the background, and the field-angle configuration is also illustrated. Clear oscillations can be observed above 15~T, showing the clean nature of the samples with minimal disorder. Details of Methods are displayed in Supplementary information.\\

\textbf{3. Results and discussion}

Fast-Fourier transforms (FFT) of the data at various temperatures for the two samples at two different field angles are shown in Figs.~\ref{FFTLK}(a) and (b) (also see Supplementary information 3 and 4 for different field angles and FFT ranges). Both when the field is applied in the $ab$-plane at $\varphi=39^{\circ}$, and  rotated out of the $ab$-plane at $\theta=19^{\circ}$ five oscillation frequencies are identified, which were assigned labels based on a comparison to DFT calculations for the paramagnetic state with the Ce-$4f$ electrons being fully localized core electrons. Figure~\ref{FFTLK}(c) displays the temperature dependence of the peak amplitude corresponding to the $\alpha^{\prime}$ frequency for sample A  and the $\gamma_2$ frequency for sample B. The dashed lines show the results from fitting the amplitudes using the temperature dependent part of the  Lifshitz-Kosevich formula \cite{Shoenberg2009,LKformulasimple}, $A=A_0(14.69m^{*}T/B_{m})/[{\rm sinh}(14.69m^{*}T/B_{m})]$, where $m^*$ is the effective carrier mass, and  $B_{m}$ is the average inverse-field of the FFT field window. The fitted $m^*$ are $2.49m_{e}$ for the $\alpha^{\prime}$ pocket and $3.9m_e$ for the $\gamma_2$ pocket, while the respective effective masses from the DFT calculations $m_{\rm calc}$ are $1.66m_{e}$ and $2.51m_e$. The moderate enhancement of the observed values($m^*/m_{\rm calc}\approx1.5$) suggests the influence of  electronic correlations. The mass enhancement is smaller than the enhancement of the Sommerfeld coefficient over the value calculated from the density of states $\gamma_{\rm calc}=11$~mJ~mol$^{-1}$~K$^{-2}$, where $\gamma/\gamma_{\rm calc}$ is 14 or 34 using $\gamma$ derived from the paramagnetic and FM states, respectively \cite{Shen2019}. This suggests that other orbits such as those on the $\beta$ branches may have significantly larger $m^*$. From the Dingle analysis, mean free paths $l_{\rm tr}$ of 53 and 304~nm are obtained for the $\alpha^{\prime}$ and $\gamma_2$ orbits, respectively, consistent with the estimates from the resistivity of  $l_{\rm tr}\approx100$~nm  (see Supplementary information 2). These respectively yield $k_{\rm F}l_{\rm tr}$ of 334 and 1916, well satisfying the clean limit criterion of $k_{\rm F}l_{\rm tr}\geq300$ for FM quantum phase transitions \cite{Sang2014}.

\begin{figure} [t]
	\includegraphics[angle=0,width=0.99\textwidth]{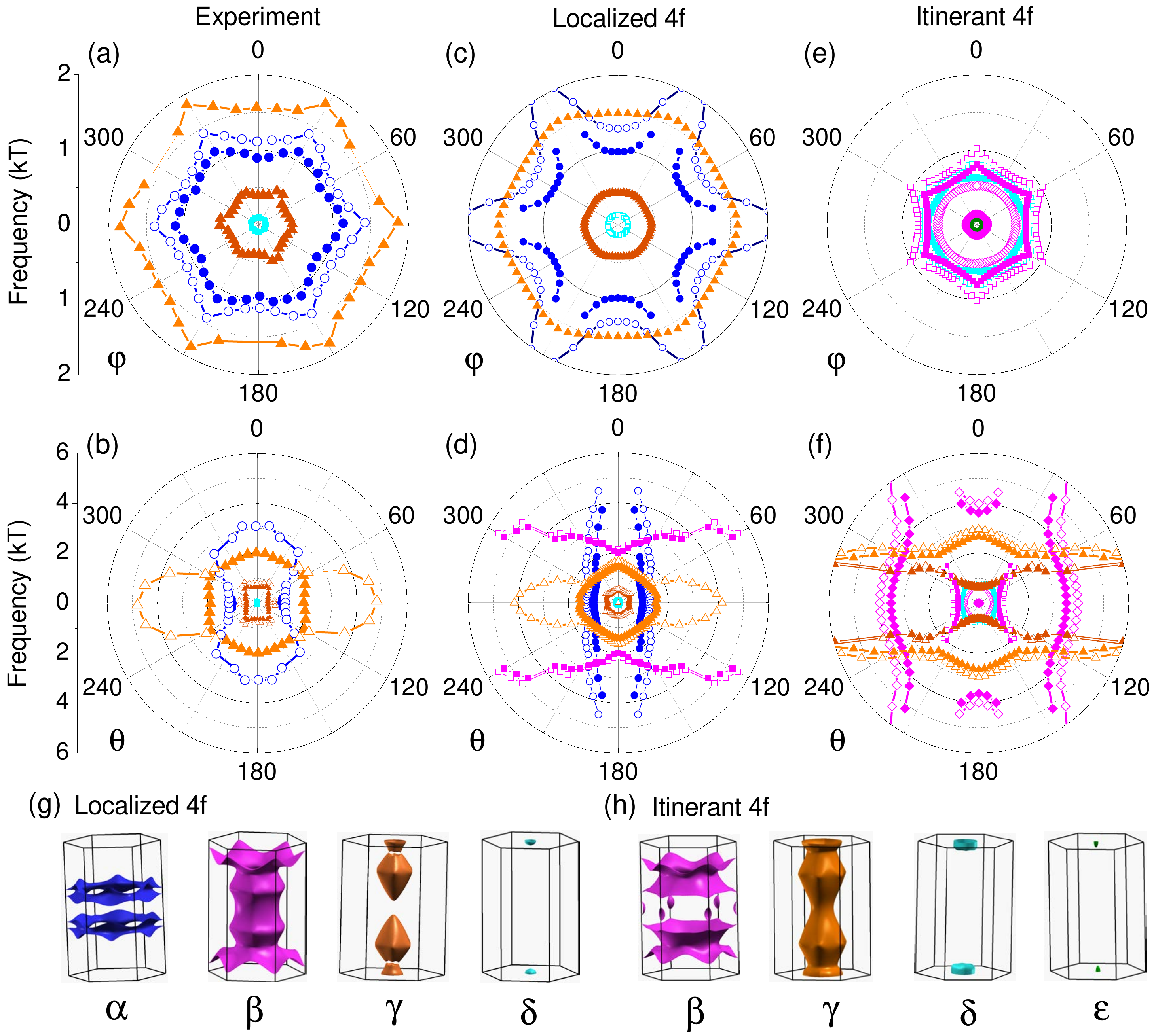}
	\centering
	\vspace{-10pt}
	\caption{\label{angle} (Color online) Experimental angle dependence of the quantum frequencies for (a) sample A using the torque method at different field angles $\varphi$ in the $ab$-plane, and (b) sample B using the tunnel-diode oscillator method at different field angles $\theta$ from the $c$-axis. $\gamma_1$, $\gamma_2$, $\gamma_1^{\prime}$ and $\gamma_2^{\prime}$ are shown by the solid dark-orange, solid light-orange, open dark-orange, and open light-orange triangles, respectively. The solid and empty blue circles correspond to $\alpha$ and  $\alpha^{\prime}$, respectively, while the $\delta$ frequencies are solid cyan triangles. The corresponding angular dependences of the quantum frequencies from calculations where the $4f$ electrons are fully localized are displayed in (c) and (d), while the case for itinerant $4f$-electrons is shown in (e) and (f). For the latter, the solid and empty magenta symbols correspond to $\beta$ pockets, while the small frequencies in dark-green are from the new $\epsilon$ surfaces. Four Fermi surfaces are displayed in (g) for the localized case, and (h) for the itinerant case. Here one of the pair of ASOC-split pairs of Fermi surfaces are displayed, where the colors match the symbols of the corresponding orbits in panels (a)-(f).}
\end{figure}

In order to map the Fermi surface, the experimental setup was rotated with respect to the field direction, in 10$^{\circ}$ intervals. The results for the two samples are displayed in Figs.~\ref{angle}(a) and (b), where the experimental data is shown across the full angular range in accordance with the lattice symmetry. Figures~\ref{angle}(c) and (d) display the calculated angular dependence of the quantum oscillation (QO) frequencies derived from DFT calculations with localized $4f$ electrons, while Figs.~\ref{angle}(e) and (f) show those from calculations with itinerant $4f$ electrons  (see Supplementary information 4 for the full frequency range). A moderately strong ASOC lifts the degeneracy of the bands ($\approx$50~meV splitting at the $K$-point), splitting each Fermi surface into a pair of sheets, and Fig.~\ref{angle}(g) displays  one of the ASOC-split Fermi surfaces for each of the four sets of  surfaces in the case of localized $4f$ electrons (see Supplementary information 5 for all surfaces). Those labelled $\alpha$ correspond to a pair of ring-like surfaces close to the  zone edge, while $\beta$ is open with extended surfaces at the top and bottom of the zone, $\gamma$ are dummy-shaped pockets separated at the zone center, and $\delta$ are very small pockets at the  $A$ point. For itinerant $4f$-electrons [Fig.~\ref{angle}(h)], the $\alpha$-surfaces are no longer found, and very small $\epsilon$-surfaces appear. In addition, the $\beta$ pocket splits into a number of small pockets together with a pair of ring-like pockets, and $\gamma$ is now connected continuously along $\Gamma-A$.

In general, there is a good correspondence between the experimental data and the calculations assuming localized $4f$ electrons.  Frequencies corresponding to orbits for three of these sets, $\alpha$,  $\gamma$ and $\delta$, can be assigned to the experimental observations, while the larger frequencies associated with the $\beta$ pockets were not resolved.  A small frequency corresponding to the $\delta$ pockets is observed, which changes little upon varying the field direction.  For fields within the $ab$-plane, a pair of frequencies are observed along all directions with values of around 1000~T, labelled $\alpha$ and $\alpha^{\prime}$, which match the  ASOC-split $\alpha$ pockets. These orbits  show a moderate anisotropy within the $ab$-plane, but a large out-of-plane anisotropy, in agreement with the calculations. Experimentally  $\alpha$ and $\alpha^{\prime}$ are observed for all $\varphi$ within the $ab$-plane, whereas from the calculations these pockets are absent at $\varphi=30^\circ$ (and other positions related by six-fold symmetry). This difference may be explained by the volume of these pockets being slightly smaller than that found from calculations, which breaks $\alpha$ and $\alpha^{\prime}$  into six isolated finite pockets, leading to closed orbits for fields along all in-plane directions.

When the $4f$ electrons are localized, one of the ASOC-split  dummy-shaped $\gamma$ surfaces does not extend down to the $A$ point, but terminates above the zone edge, and there is a second smaller pocket around the $A$ point [as shown in Fig.~\ref{angle}(g)]. Consequently, for all field directions there are two frequencies associated with this surface, one for the smaller pocket ($\gamma_1\approx500$~T, $H\parallel ab$), and another for the larger one ($\gamma_2\approx1600$~T, $H\parallel ab$). For the other ASOC-split surface, the  pocket remains fully connected down to the $A$ point (see Supplementary information 5), and therefore for $ab$-plane fields there is only expected to be a single frequency $\gamma_2^{\prime}\approx5000$~T, which is considerably larger than $\gamma_2$. On the other hand, for fields close to the $c$-axis, an additional frequency $\gamma_1^{\prime}$ is expected, slightly larger than $\gamma_1$.

It can be seen from Figs~\ref{angle}(a)-(d), that there is reasonably good agreement between the observed and calculated angular dependences of the $\gamma$ pockets. In Fig.~\ref{angle}(a), the two smaller frequencies associated with one surface, $\gamma_1$ and $\gamma_2$ can be detected, but the larger $\gamma_2^{\prime}$ frequency is not.  $\gamma_2^{\prime}$ can however be seen in the TDO measurements in Figs.~\ref{angle}(b), and this difference is likely due to the greater sensitivity of the TDO method to higher oscillation frequencies. In addition, $\gamma_2$ could not be detected along a small angular range for fields in the $ab$-plane (around $\varphi=60^{\circ}, 180^{\circ}, 300^{\circ}$). We note that the quantum oscillations were particularly weak for measurements with these field directions (see Supplementary information 4), which may be due to the Fermi surface geometry along certain directions \cite{Shoenberg2009}. Note that for fields close to the $c$-axis we observe a frequency just above  $\gamma_1$, which is consistent with $\gamma_1^{\prime}$. However, this frequency can still be detected for $ab$-plane fields, contrary to the calculations.

On the other hand, the angular dependence of the QO frequencies assuming the $4f$ electrons to be itinerant [Figs.~\ref{angle}(e) and (f)]  are drastically different from those observed experimentally. In particular, since the $\alpha$ pockets are absent, and $\gamma$ are continuously connected along $\Gamma-A$, for fields in the $ab$-plane, all the predicted QO frequencies are relatively small ($<800$~T), besides two-large frequencies associated with the ASOC-split $\beta$ surfaces ($>3500$~T). This is in stark contrast to the data in Fig.~\ref{angle}(a), where three frequencies ($\gamma_2$, $\alpha$ and $\alpha^{\prime}$) are observed in the intermediate range 1000-2000~T. Furthermore, the calculations upon varying $\theta$ for the itinerant case are also very different from the data, largely as a consequence of the $\beta$ and $\gamma$ surfaces being open along significant angular regions, in contrast to observations.

Our findings that the electronic structure of CeRh$_6$Ge$_4$ is largely consistent with the scenario that the Ce-$4f$ electrons remain localized strongly resembles the case of CeRhIn$_5$, where quantum oscillation studies at ambient pressure reveal that the  Ce-$4f$ electrons do not contribute to the Fermi surface, in stark contrast to isostructural CeCoIn$_5$ and CeIrIn$_5$ where the $4f$  electrons have an itinerant nature \cite{Shishido2002,Harrison2004,Jiao2015}. Furthermore, it was found that the Fermi surface of  CeRhIn$_5$ undergoes a reconstruction upon crossing the pressure-induced QCP, together with divergent cyclotron masses \cite{Shishido2005}. These observations are ascribed to the delocalization of the $4f$ electrons at a local-type QCP \cite{Si2010}. In CeRhIn$_5$, the identification of local quantum criticality is corroborated by strange metallic behavior in the resistivity, and a divergent specific heat coefficient reaching a very large low temperature value ($>1$~J~Mol$^{-1}$K$^{-2}$) \cite{Park2006,Park2009}. These are strikingly similar to the resistivity and specific heat of CeRh$_6$Ge$_4$ at the FM QCP \cite{Shen2019}, but final confirmation of local quantum criticality requires probing the electronic structure  of CeRh$_6$Ge$_4$  under pressure.

We note that there are some minor differences between the experimental results and calculations with localized $4f$-electrons, namely  that  the   hole pockets $\alpha$ are found to be enclosed experimentally, while they are open  in calculations, and  the observation of the $\gamma_1^{\prime}$ frequency in the $ab$-plane. In a real material, some hybridization between Ce-4$f$ electrons and conduction electrons is inevitable, which is not taken into account by calculations where all the $f$ electrons are fully localized as core electrons. Should the hybridization correctly account for the size of the hole pockets, this could resolve this discrepancy. Note that a moderate influence of hybridization between the $4f$ and conduction electrons on the band structure of CeRh$_6$Ge$_4$ and CeRhIn$_5$ was also inferred from angle-resolved photoemission spectroscopy measurements \cite{Chen2018}. Moreover, in CeRhIn$_5$ there are also some discrepancies between the observed QO frequencies and DFT calculations with localized $4f$ electrons \cite{Jiao2015}. The origin of the discrepancies remains an open question, and further studies are necessary to determine whether this may be accounted for by hybridization effects. On the other hand, DFT calculations with itinerant $4f$ electrons are vastly different to the quantum oscillation results, supporting  the localized nature of the $4f$-shell. Moreover, this conclusion is corroborated by calculations using the LDA+U method (see Supplementary information 5), which are very close to those with fully localized $4f$ electrons.\\

\textbf{4. Conclusion}

In summary, our observation of the quantum oscillation in CeRh$_6$Ge$_4$ shows  that the FM QCP occurs in a clean system. The similarity between the quantum oscillation frequencies and calculations with the  $4f$ electrons as core electrons provides evidence for the Ce-$4f$ electrons remaining localized at ambient pressure, and not contributing to the Fermi surface. These findings indicate that the electronic structure of  CeRh$_6$Ge$_4$ is distinct from itinerant ferromagnets where quantum criticality is avoided \cite{Brando2016}, and  suggests that localized magnetism is crucial for the appearance of a FM QCP. On the other hand, in order to determine the nature of the QCP,  it is highly demanded to probe the evolution of the electronic structure of CeRh$_6$Ge$_4$ under pressure, by measurements such as the quantum oscillation or Hall resistivity, so as to look for the predicted jump of the Fermi surface volume upon crossing the QCP  \cite{Shen2019}.\\

\textbf{Conflict of interest}

The authors declare that they have no conflict of interest.

\textbf{Acknowledgements}

We acknowledge valuable discussions with John Singleton. This work was supported by the National Key R\&D Program of China (No.~2017YFA0303100, No.~2016YFA0300202), the National Natural Science Foundation of China (No.~12034017, No.~U1632275 and No.~11974306), and the Science Challenge Project of China (No.~TZ2016004). A portion of this work was performed at the National High Magnetic Field Laboratory, which is supported by the National Science Foundation Cooperative Agreement No. DMR-1644779 and the State of Florida.

\textbf{Author contributions}
The project was conceived by Huiqiu Yuan. The crystals were grown by Yongjun Zhang; quantum oscillation measurements including cantilever and TDO were performed by An Wang and David Graf and analyzed by An Wang, Bin Shen, Ye Chen, Feng Du, Yang Liu, Michael Smidman and Huiqiu Yuan; resistivity and heat capacity measurements were performed by Yongjun Zhang and Bin Shen; band structure calculation was performed by Feng Du and Chao Cao; An Wang, Feng Du, Chao Cao, Michael Smidman, Frank Steglich and Huiqiu Yuan wrote the manuscript with inputs and from all authors.


\begin{figure} [tb]
	\includegraphics[angle=0,width=0.7\textwidth]{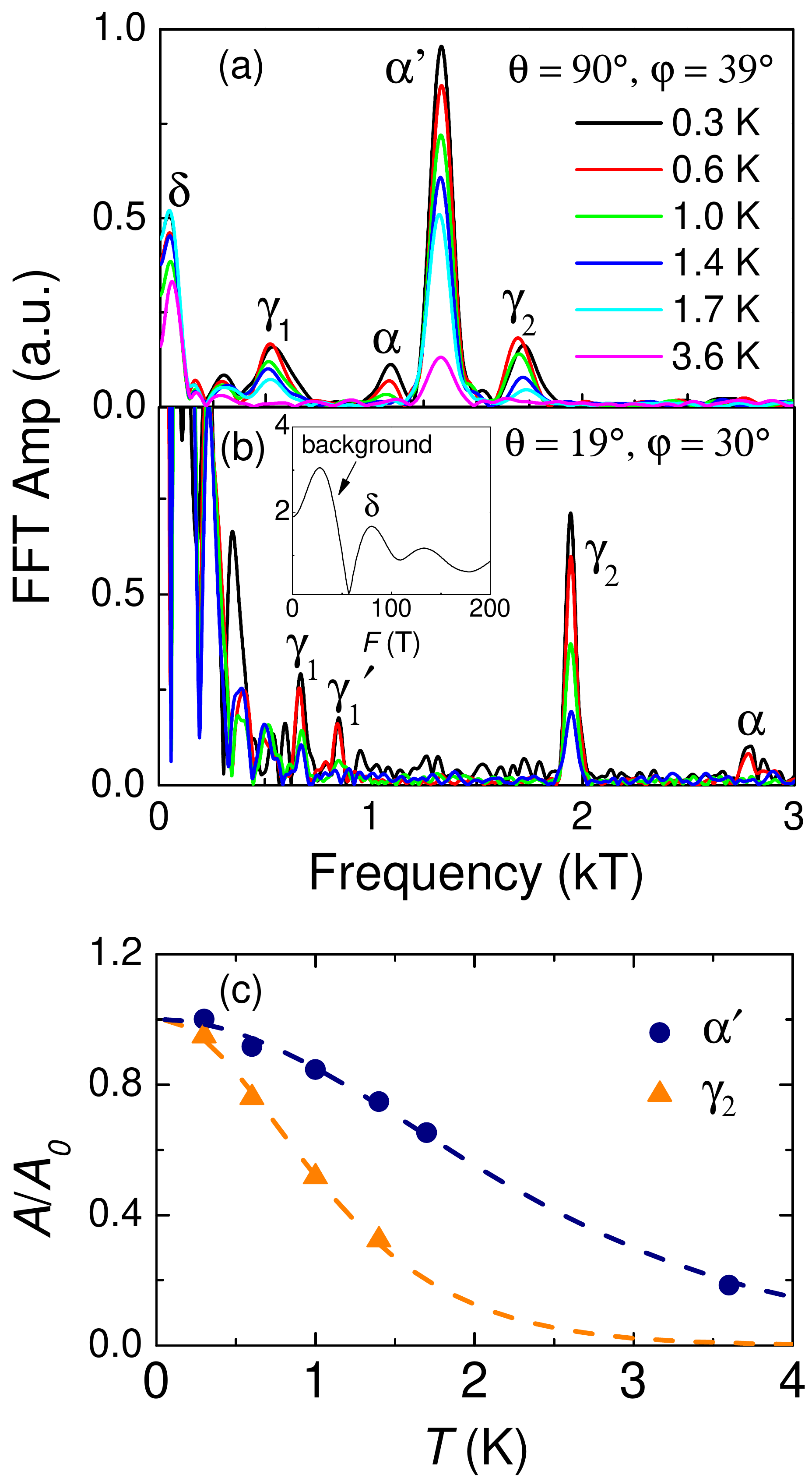}
	\centering
	\vspace{-10pt}
	\caption{\label{FFTLK} (Color online) FFT of the quantum oscillations at various temperatures are dislayed for (a) sample A with an FFT field window of  $25-45~$T, and (b)  sample B with a field window of $15-35$~T. The inset shows the low frequency region, where the $\delta$ peak is labelled, while the nearly temperature independent peak at lower frequencies is likely  due to vibrations of the magnet. (c) Temperature dependence of the FFT amplitudes of  the frequencies corresponding to $\alpha^{\prime}$ in (a) (for $35-40$~T), and $\gamma_2$ in (b) (for $25-30~$T). The dashed lines show the results from fitting using the Lifshitz-Kosevich formula. }
\end{figure}

\end{document}